\documentclass[aps,twocolumn,showpacs,amsmath,amssymb,floatfix]{revtex4-1}
\usepackage[dvips]{graphicx}
\usepackage[colorlinks=true]{hyperref}
\usepackage{bm}
\usepackage{color}
\usepackage{ulem}

\usepackage[english]{babel}
\addto{\captionsenglish}{}

\begin{document}

\title{Ground-state phase diagram of the triangular lattice Hubbard model by the density-matrix renormalization group method}

 \author{Tomonori Shirakawa$^{1}$} 
\author{Takami Tohyama$^{2}$}
\author{Jure Kokalj$^{3,4}$}
\author{Sigetoshi Sota$^{5}$}
\author{Seiji Yunoki$^{1,5,6}$}
\affiliation{
${}^1$Computational Quantum Matter Research Team, RIKEN Center for Emergent Matter Science (CEMS), Wako, Saitama 351-0198, Japan \\
${}^2$Department of Applied Physics, Tokyo University of Science, Tokyo 125-8585, Japan \\
${}^3$Jo\v{z}ef Stefan Institute, Jamova c. 39, 1000 Ljubljana, Slovenia  \\
${}^4$Faculty of Civil and Geodetic Engineering, University of Ljubljana, SI-1000 Ljubljana, Slovenia \\
${}^5$Computational Materials Science Research Team, RIKEN Advanced Institute for Computational Science (AICS), Kobe, Hyogo 650-0047, Japan \\
${}^6$Computational Condensed Matter Physics Laboratory, RIKEN, Wako, Saitama 351-0198, Japan
}

\date{\today}

\begin{abstract}
Two-dimensional density-matrix renormalization group method is employed to examine the ground 
state phase diagram of the Hubbard model on the triangular lattice at half-filling. 
The calculation reveals two discontinuities in the double occupancy with increasing 
the repulsive Hubbard interaction $U$ at 
$U_{\rm c1}\sim 7.8t$ 
and $U_{\rm c2}\sim 9.9t$ 
($t$ being the hopping integral), 
indicating that there are three phases separated by first order transitions. 
The absence of any singularity in physical quantities for $0 \le U < U_{\rm c1}$ 
implies a metallic phase in this regime. 
For $U > U_{\rm c2}$, the local spin density induced by an applied pinning magnetic field 
exhibits a three sublattice feature, which is compatible with the $120^{\circ}$ N\'eel ordered state 
realized in the limit of $U \to \infty$. 
For $U_{\rm c1} < U < U_{\rm c2}$, a response to the applied pinning magnetic field is comparable to 
that in the metallic phase with a relatively large spin correlation length, but showing 
neither valence bond nor chiral magnetic order, 
which therefore resembles gapless spin liquid. 
However, the spin structure factor for the intermediate phase exhibits the maximum 
at the ${\rm K}$ and ${\rm K}^{\prime}$ points in the momentum space, 
which is not compatible to spin liquid with a large spinon Fermi surface. 
The calculation also finds that the pairing correlation function monotonically decreases 
with increasing $U$ and thus the superconductivity is unlikely in the 
intermediate phase. 
\end{abstract}

\maketitle

\section{Introduction}

There have been accumulating experimental evidences that 
several organic materials, $\kappa$-(BEDT-TTF)$_2$Cu$_2$(CN)$_3$~\cite{shimizu}, 
EtMe$_3$Sb[Pd(dmit)$_2$]$_2$~\cite{itou, yamashita1, yamashita2}, and 
$\kappa$-H$_3$(Cat-EDT-TTF)$_2$~\cite{isono}, forming a quasi two-dimensional (2D) triangular 
structure and exhibit quantum spin liquid 
(QSL)~\cite{patricklee,balentz}, where any spatial symmetry breaking does not occur 
due to the quantum fluctuation, even when it is cooled down to zero temperature. 
The realization of QSL against a symmetry-broken ordered state 
in higher spatial dimensions more than one dimension 
is one of the long standing issues in condensed matter physics~\cite{patricklee} since the first 
proposal of resonating valence bonds (RVB) states by Anderson~\cite{anderson}. 
It has been considered that one of the key ingredients for the emergence of stable QSL is geometrical 
frustration~\cite{anderson}, which increases quantum fluctuations and thus prevents 
symmetry breaking. 
In this context, the spin-1/2 antiferromagnetic Heisenberg model on the triangular lattice had 
been considered~\cite{fazekas}. 
However, recent numerical studies, including two-dimensional density-matrix renormalization 
group (2D-DMRG) analysis, have suggested that the ground state of the spatially isotropic 
model is $120^{\circ}$ N\'eel ordered~\cite{bernu, capriotti, zheng, white3}.

In addition to the geometrical frustration, other factors for stabilizing QSL have also been considered, 
such as i) the spatially anisotropic exchange interactions~\cite{yunoki}, ii) the higher order corrections 
of exchange interactions~\cite{motrunich}, and iii) the charge degree of freedom~\cite{note,hotta,naka,watanabe2017}. 
The later two are captured by the triangular lattice Hubbard model at half electron filling described 
by the Hamiltonian  
\begin{eqnarray}
\mathcal{H} = - t \sum_{\left< i,j \right>} \sum_{\sigma = \uparrow,\downarrow} 
( c_{i,\sigma}^{\dagger} c_{j,\sigma} + {\rm h.c.} ) + U \sum_i n_{i,\uparrow} n_{i,\downarrow}, 
\label{eq:ham}
\end{eqnarray}
where $c_{i,\sigma}$ ($c_{i,\sigma}^{\dagger}$) represents the annihilation (creation) operator 
of an electron with spin $\sigma\, (= \uparrow,\downarrow)$ at site $i$ on the triangular lattice, 
$n_{i,\sigma}= c_{i,\sigma}^{\dagger} c_{i,\sigma}$, 
and the sum $\left< i, j \right>$ runs over all pairs of nearest-neighbor sites $i$ and $j$. 
Indeed, the QSL phase in the organic materials appears next to the metallic phase, 
indicating that the QSL occurs close to the Mott metal-insulator 
transition~\cite{kurosaki, furukawa, kandpal} where the above two factors ii) and iii) 
are important. 
In fact, it has been extensively argued that the triangular lattice Hubbard model is the 
simplest effective model 
to describe and understand the metal-insulator transition and 
the QSL phase in the organic materials~\cite{powell}.

Elucidating the ground state phase diagram of the triangular lattice Hubbard model 
at half-filling is 
a challenge for numerical techniques in strongly correlated electron systems. 
Various numerical methods~\cite{koretsune, clay, kokalj, sahebsara, yamada, morita, yoshioka, yang, 
kyung, liebsch, galanakis, ohashi, sato, lee, dang, li, watanabe1, watanabe2, tocchio1, tocchio2} 
have been applied so far, but the results are, nevertheless, controversial. 
The exact diagonalization techniques~\cite{koretsune, clay, kokalj}, the variational cluster 
approximation (VCA)~\cite{sahebsara, yamada}, 
the path-integral renormalization group (PIRG) method~\cite{morita, yoshioka}, 
as well as the recently proposed ladder dual-Fermion approach~\cite{li} have suggested that 
there exist three phases in the ground state phase diagram with increasing $U/t$, 
i.e., a metallic phase, a nonmagnetic insulating phase, and the $120^{\circ}$ N\'eel ordered 
insulating phase. 
This is also supported by the numerical analysis of an effective strong coupling spin model~\cite{yang}. 
On the other hand, the variational Monte Carlo 
methods~\cite{watanabe1, watanabe2, tocchio1, tocchio2} have suggested the absence of the 
nonmagnetic insulating phase. Besides the presence or absence of the intermediate insulating 
phase, the critical $U_c/t$ for the metal-insulator transition significantly 
varies among different methods~\cite{kokalj}.

Recently, the 2D-DMRG method has been applied to various 2D strongly correlated quantum systems
~\cite{stoudenmire, jiang1, gong1, yan, jiang2, dependbrock, nishimoto, gong2, gong3, ganesh1, ganesh2, shinjo, tohyama, zhu, hu, sellmann, shinjo2, okubo2017}, 
although the DMRG method is best performed for one-dimensional gapful 
systems~\cite{white1,white2,schollwock,hallberg1}. This is because the 2D-DMRG calculations 
with keeping large enough number of adapted density-matrix eigenstates to guarantee 
the desired numerical accuracy have become possible within reasonable computational resources, 
specially, for 2D spin-1/2 Heisenberg models~\cite{stoudenmire}.

Here, we employ the 2D-DMRG method to examine the ground state phase diagram of 
the repulsive Hubbard model on the triangular lattice at half electron filling. 
Our calculation reveals two discontinuities in the double occupancy of electrons 
with increasing $U/t$ at $U_{\rm c1}= 7.55t\sim8.05t$
and $U_{\rm c2}= 9.65t\sim10.15t$ 
for three different clusters up to 48 sites, strongly indicating that there are three phases 
separated by first order transitions at $U_{\rm c1}$ and $U_{\rm c2}$. 
The spin oscillation pattern for $U > U_{\rm c2}$ under a pinning magnetic field 
exhibits a three sublattice feature, 
compatible with the $120^{\circ}$ N\'eel ordered state. 
Moreover, the spatial distribution of 
the nearest-neighbor spin correlation is found to be quite different among the three 
phases. The suppression of oscillatory behavior in the intermediate phase at 
$U_{\rm c1} < U < U_{\rm c2}$ suggests 
this phase in neither bond order nor valence bond solid.
In addition, the spin correlation length 
in the intermediate phase is found to be 
larger than that for $U < U_{\rm c1}$ but smaller than that for $U > U_{\rm c2}$. 
Furthermore, the response to a pinning magnetic field in the intermediate phase 
is rather comparable to that in the paramagnetic metallic state.
These features in the intermediate phase resembles gapless spin liquid~\cite{powell, normand}. 
However, the spin structure factor in the intermediate phase shows 
a single maximum at the ${\rm K}$ and ${\rm K}^{\prime}$ points in the momentum space, 
which is not compatible with the expectation for the spinon Fermi sea state~\cite{motrunich}. 
Superconductivity is also excluded in the intermediate phase.

The rest of this paper is organized as follows. 
First, the shape of 2D clusters studied here is introduced and the convergence of the DMRG calculations 
is discussed in Sec.~\ref{sec:method}. 
Section~\ref{sec:cal} is devoted to our results for the triangular lattice Hubbard model at half-filling. 
We first show the ground state energy and the double occupancy to reveal the existence of three phases 
in Sec.~\ref{sec:energy}. Next, we explore the properties of the ground state 
in each phase by calculating different quantities, including the response to a pinning magnetic 
field in Sec.~\ref{sec:pinning}, the spin correlation function in Sec.~\ref{sec:spincorrelation}, 
the spin structure factor in Sec.~\ref{sec:sq}, 
the spatial distribution of the nearest-neighbor spin and bond correlations in Sec.~\ref{sec:nn_spin} and 
Sec.~\ref{sec:bond}, respectively, the chiral correlation function in Sec.~\ref{sec:chiral}, 
and the pairing correlation function in Sec.~\ref{sec:scpair}. 
We then discuss possible relevance to the experimental observation and provide 
several remarks in Sec.~\ref{sec:discussion}, before summarizing the paper in Sec.~\ref{sec:summary}. 
In Appendix~\ref{app:egap}, we examine the entanglement gap of the ground state as a function of $U/t$.

\section{\label{sec:method}Method}

\begin{figure}[htbp]
\begin{center}
\includegraphics[width=\hsize]{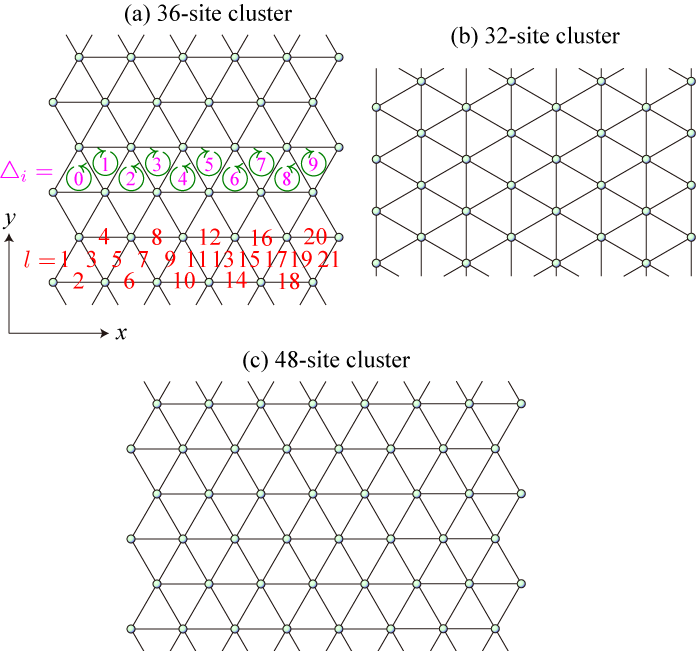}
\caption{The (a) 36-, (b) 32-, and (c) 48-site clusters. Open (periodic) boundary conditions are 
imposed in the $x$ direction ($y$ direction). The indexing of bonds ($l=1,2,3,\cdots,21$) as well as 
elementary triangles (${\triangle_i}=0,1,2,...,9$) with their chiral directions (arrows)
are indicated in (a).
}
\label{cylinder}
\end{center}
\end{figure}

We consider 32-, 36- and 48-site clusters depicted in Fig.~\ref{cylinder}.
Since the results for these different clusters are qualitatively the same, 
we shall mainly show the results for the 36-site cluster. 
Following the notation in Refs.~\onlinecite{zhu,hu}, 
clusters forming the triangular lattice can be classified as XC$n$ (YC$n$) 
where the bond direction of a cluster 
is parallel to the $x$ direction ($y$ direction), as shown in Fig.~\ref{cylinder}, 
and $n$ in XC$n$ (YC$n$) represents 
the number of bonds in zigzag (vertical) $y$ direction. 
Accordingly, the 36- and 48-site clusters belong to XC6, and the 32-site cluster belongs to YC4.

\begin{figure}[htbp]
\begin{center}
\includegraphics[width=\hsize]{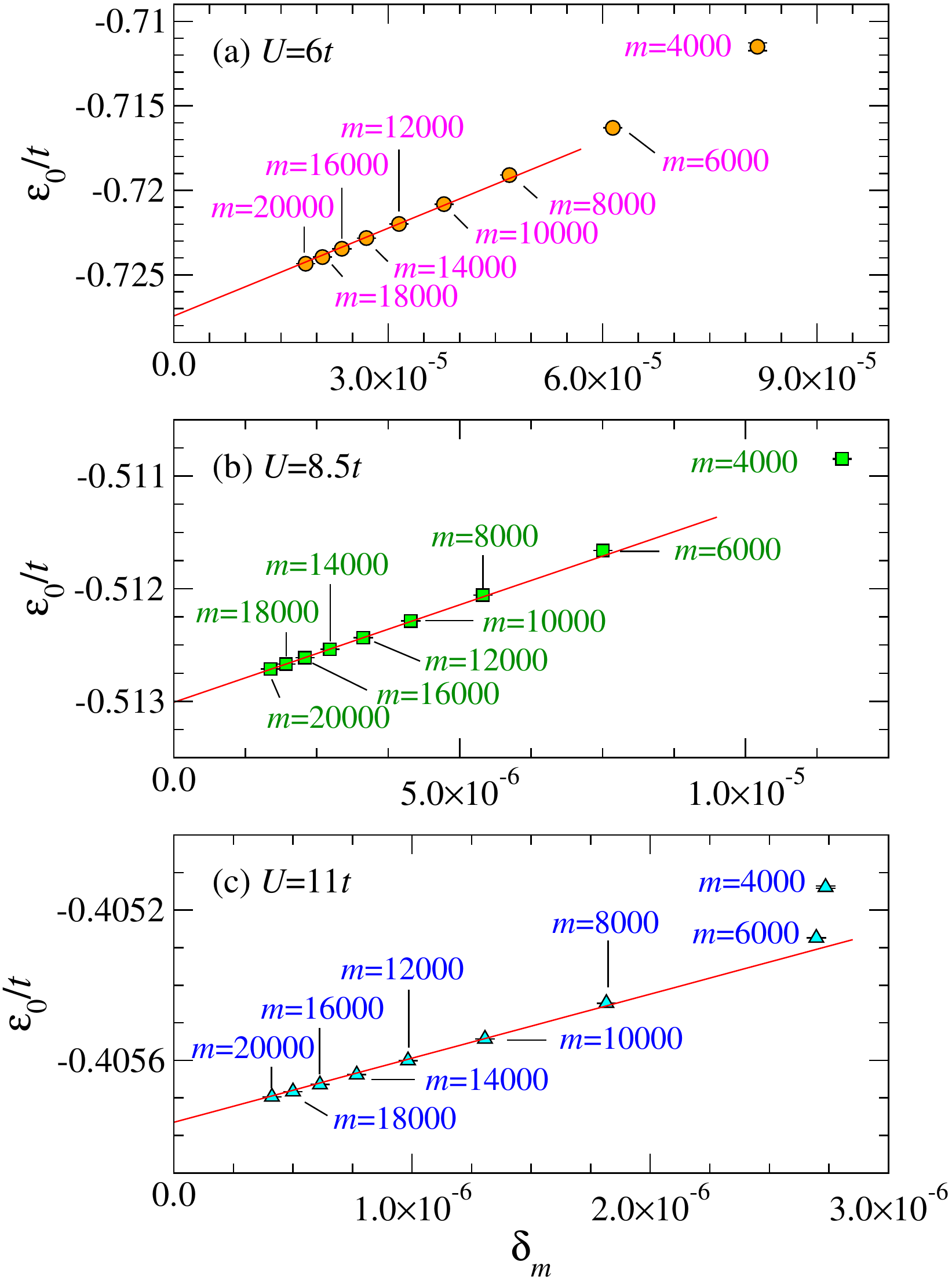}
\caption{(color online) Ground state energy per site $\varepsilon_0$ 
as a function of the discarded weight $\delta_m$ for (a) $U=6t$, (b) $U=8.5t$, and (c) $U=11t$. 
The cluster used here is the 36-site cluster shown in Fig.~\ref{cylinder}(a). 
The number $m$ of eigenstates of the reduced density-matrix kept in the DMRG calculations 
is indicated beside each data point. 
A red straight line shows a linear fit to the three data points with $m=10\ 000$, $12\ 000$, and $14\ 000$. }
\label{extrapenergy}
\end{center}
\end{figure}

Figure~\ref{extrapenergy} shows the convergence of the ground state energy for the 36-site cluster, 
as a function of the discarded weight $\delta_m$ defined as
\begin{eqnarray}
\delta_m = 1 - \sum_{n=1}^m \lambda_{n}, 
\end{eqnarray} 
where $\lambda_{n}$ is the $n$th largest eigenvalue of the reduced density-matrix of the ground state. 
As shown in Fig.~\ref{extrapenergy}, we find that the ground state energies for $m \geq 10\ 000$ 
scale linearly with $\delta_m$, implying that the convergence of our calculations is well controlled.

Throughout the study, we set the $z$ component of total spin to be zero.
We keep up to $m=10\ 000$ density-matrix eigenstates for the 32-site cluster, 
$m=14\ 000$ for the 36-site cluster, and 
$m=20\ 000$ for the 48-site cluster. 
As shown in Fig.~\ref{extrapenergy}, when we use $m=14\ 000$ density-matrix eigenstates for the 36-site cluster, 
the typical orders of the discarded weight 
are $ 2.7 \times 10^{-5}$ for $U=6t$, $ 2.7 \times 10^{-6}$ for $U=8.5t$, and $ 7.6 \times 10^{-7}$ for 
$U=11t$. 
On the other hand, when we use $m=20\ 000$ density-matrix eigenstates for the 48-site cluster, 
the typical orders of the discarded weight are $2.6 \times 10^{-5}$ for $U=6t$, 
$2.3 \times 10^{-6}$ for $U=8.5t$, and $2.8 \times 10^{-7}$ for $U=11t$, thus obtaining 
the convergence similar to that for the 36-site cluster.

\section{\label{sec:cal}Results}

\subsection{\label{sec:energy}Energy and double occupancy}

\begin{figure}[tbhp]
\begin{center}
\includegraphics[width=0.95\hsize]{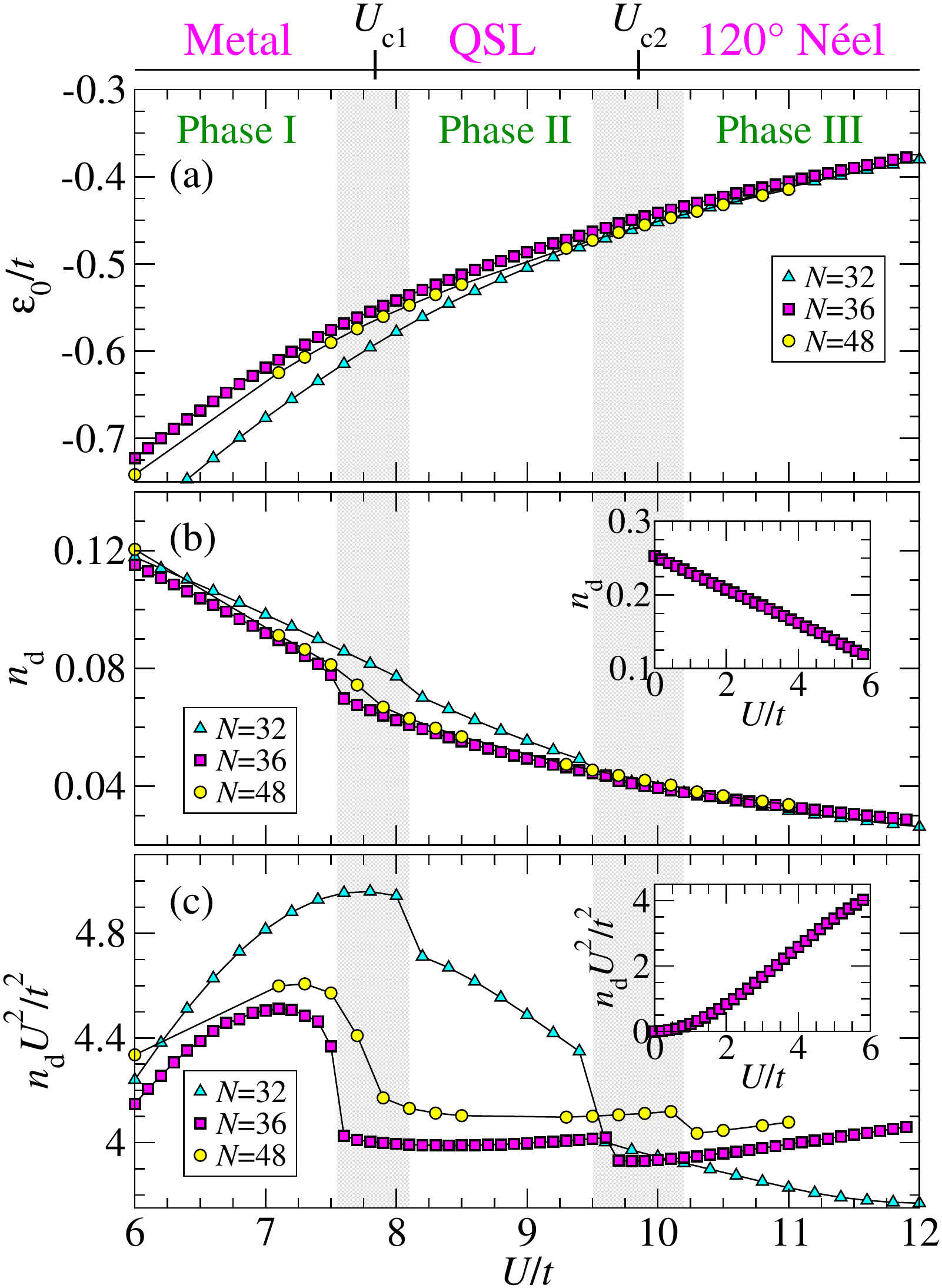}
\caption{(color online) (a) Ground state energy per site $\epsilon_0$, 
(b) double occupancy $n_d$, and (c) $n_d U^2$ for three different clusters. 
Insets in (b) and (c) show the results for small values of $U$ calculated in the 36-site cluster.
The ground state phase diagram is shown schematically 
in the top panel, where QSL denotes quantum spin liquid. 
The phase boundaries are indicated by gray shades. }
\label{phase}
\end{center}
\end{figure}

We first study the $U/t$ dependence of the ground-state energy and double occupancy. 
Figures~\ref{phase}(a) and ~\ref{phase}(b) show the ground-state energy per site, 
\begin{eqnarray}
\varepsilon_0=\langle\psi_0|\mathcal{H}|\psi_0\rangle/N
\end{eqnarray}
and the site average of the double occupancy, 
\begin{eqnarray}
n_d = \frac{1}{N}\sum_i \left< \psi_0 \right| n_{i,\uparrow} n_{i,\downarrow} \left| \psi_0 \right>,
\end{eqnarray}
where $\left| \psi_0 \right>$ is the ground-state obtained by the 2D-DMRG calculation and  
$N$ is the number of sites. 
As shown in Fig.~\ref{phase}(b), there exist two discontinuities in the double occupancy. 
It should be noted that $\varepsilon_0$ and $n_d$ are related via
$n_d=\partial\varepsilon_0/\partial U$. 
We have numerically verified this relation, 
supporting the satisfactory convergence of our results.

As shown in Fig.~\ref{phase}(c), the discontinuities in the double occupancy are most apparent 
when $n_d U^2$ is plotted. 
The first discontinuity occurs at $U_{\rm c1}/t=7.55$-$8.05$ and the 
second one is located at $U_{\rm c2}/t=9.65$-$10.15$. 
We find that these discontinuities in $n_d U^2$ become sharper with increasing $m$, 
indicating the nature of the first-order transition. 
We also find in the insets of Figs.~\ref{phase}(b) and \ref{phase}(c) that there is no additional 
discontinuity for $0\le U < U_{\rm c1}$. 
Therefore, we conclude that there exist two first order transitions separating three phases. 
In the following, we call the three regions phases I, II, and III, as indicated in Fig.~\ref{phase}.

Let us now compare our results with the previous studies. 
Since it includes the noninteracting limit with $U=0$, phase I is regarded as the metallic phase. 
The exact diagonalization analysis of a 16-site cluster using a finite-temperature Lanczos method  
has found that the metal-insulator transition occurs at $U_{\rm c}/t = 7.5  \pm 0.5 $~\cite{kokalj}. 
The metal-insulator transition is also found at $U_{\rm c}/t \sim 7.4  \pm 0.1$ for clusters up to 36 
sites by the PIRG method~\cite{yoshioka}. These $U_{\rm c}$ values are rather similar 
to $U_{\rm c1}$ in our calculations. 
On the other hand, the metal-insulator transition found by the VCA 
is at $U_{\rm c}/t \sim 6.3$-$6.7$~\cite{sahebsara, yamada}, 
which is slightly smaller than $U_{\rm c1}$. This is probably due to smaller clusters used in 
these VCA calculations, which tend to enhance an insulating phase. 
We calculate in Appendix~\ref{app:egap} the entanglement spectrum of the ground state as a function of 
$U/t$ and find an abrupt increase of the entanglement gap in the charge sector, 
supporting that the transition between phases I and II can be regarded 
as the metal-insulator transition.

The analysis based on the strong coupling expansion of the triangular lattice Hubbard model 
for clusters up to 36 sites~\cite{yang} finds that the phase transition 
from the $120^{\circ}$ N\'eel-ordered phase to an insulating QSL phase occurs at $U_{\rm c} \sim 10 t$, 
which is close to $U_{\rm c2}$ obtained in our calculations. 
The intermediate insulating phase with $U_{\rm c2}=9.2t\pm0.3t$ is also reported in the PIRG calculations~\cite{yoshioka}. 
Based on the comparison with these previous studies, phases II and III found in our calculations 
should correspond to a QSL phase and the $120^{\circ}$ N\'eel-ordered phase, respectively. 
In the following, we shall examine the nature of these phases.

\subsection{\label{sec:pinning}Response to a pinning magnetic field}

\begin{figure}[tbhp]
\begin{center}
\includegraphics[width=0.85\hsize]{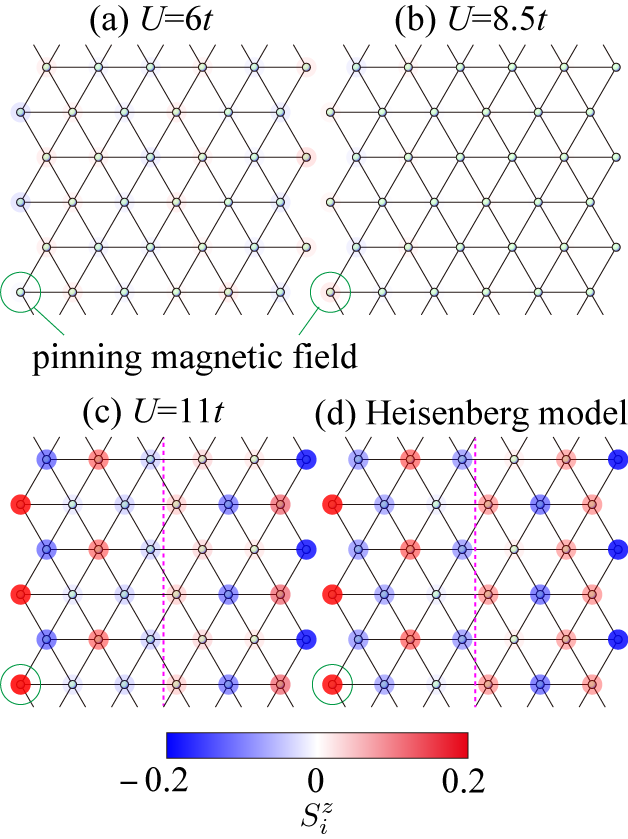}
\caption{(color online) Local spin density $S^z_i$ for (a) $U=6t$, (b) $U=8.5t$, and (c) $U=11t$
when a pinning magnetic field $h = 0.005 t$ is applied along the $z$ direction at a single site 
located at the edge of the 36-site cluster (indicated by a green open circle). 
For comparison, the results for the spin-1/2 antiferromagnetic Heisenberg model on the triangular lattice 
with the nearest-neighbor exchange interaction $J$ is also shown in (d). 
Here, the same 36-site cluster is used as in (a)--(c) and the applied pinning magnetic field is 
$h=0.005J$.
The dashed line in (c) and (d) indicates a domain wall separating the cluster into two pieces, each of which exhibits 
a three sublattice pattern, expected for the $120^{\circ}$ N\'eel order. Note that 
the $z$ component of total spin is kept zero, i.e., $\sum_{i=1}^NS_i^z=0$, in the calculations. }
\label{pinning}
\end{center}
\end{figure}

Let us first explore a possible magnetic order by applying a pinning magnetic field 
along the $z$-direction at a single site located at the edge of the cluster (see Fig.~\ref{pinning}). 
The pinning magnetic field applied at site $i_{\rm imp}$ is described by the following Hamiltonian: 
\begin{eqnarray}
\mathcal{H}^{\prime} = -\sum_i h_i\mathcal{S}_i^z, 
\end{eqnarray}
where $h_i=h\delta_{i,i_{\rm imp}}$, $\delta_{i,j}$ is the Kronecker delta 
(i.e., $\delta_{i,j}=1$ only when $i=j$), 
and $\mathcal{S}_i^z = \frac{1}{2}(n_{i,\uparrow} - n_{i,\downarrow})$. 
The results of the local spin density 
\begin{equation}
S_i^z = \langle \psi_0 | \mathcal{S}_i^z | \psi_0 \rangle 
\end{equation}
are summarized in Fig.~\ref{pinning}. 
Note that the local spin density $S_i^z$ is zero in the absence of the pinning magnetic field. 

As shown in Fig.~\ref{pinning}(c), the local spin density for $U=11t$ in phase III exhibits 
a three sublattice pattern, compatible with the $120^{\circ}$ N\'eel order, 
except that there exists a domain wall at the center of the cluster running along the $y$ direction. 
Indeed, the spatial distribution of the local spin density found here, including the domain wall structure, 
is essentially identical to that 
for the spin-1/2 antiferromagnetic Heisenberg model on the triangular lattice with the nearest-neighbor 
exchange interaction [see Fig.~\ref{pinning} (d)], 
where the ground state is 120$^{\circ}$ N\'eel ordered~\cite{bernu, capriotti, zheng, white3}. 
Therefore, this is strong evidence that the ground state in phase III is also 120$^{\circ}$ N\'eel ordered. 
In contrast, as shown in Figs.~\ref{pinning}(a) and \ref{pinning}(b), the local spin densities for 
$U=6t$ in phase I and $U=8.5t$ in phase II 
are less affected by the pinning magnetic field, 
indicating the absence of long-range magnetic order.

Applying the perturbation theory, 
the leading correction of the spin density $\Delta S_i^z $ at site $i$ is 
\begin{eqnarray}
\Delta S_i^z  \sim -h \sum_{n\,(\ne0)} 
\frac{\left< \psi_0 \right| \mathcal{S}_{i_{\rm imp}}^z \left| \psi_n \right> \left< \psi_n \right| \mathcal{S}_i^z \left| \psi_0 \right> }{E_0 - E_n}, 
\label{eq:pinning}
\end{eqnarray}
where $\left| \psi_n \right>$ is the $n$th eigenstate of $\mathcal{H}$ (without the pinning magnetic field) 
with its eigenvalue $E_n$. 
Since $\mathcal{H}$ commutes with the total spin operator, the total spin $S_{\rm tot}$ is 
a good quantum number. 
We now assume that the ground state $\left| \psi_0 \right>$ is spin singlet with $S_{\rm tot} = 0$. 
Then, the Wigner-Eckert theorem states that the matrix elements in the numerator of 
Eq.~(\ref{eq:pinning}) satisfy  
\begin{eqnarray}
\langle\psi_n|\mathcal{S}_i^z|\psi_0\rangle = \left\{ 
\begin{array}{cc}
0 & {\rm if}\ |\psi_n\rangle \not\in S_{\rm tot}=1 \\
{\rm finite} & {\rm if}\ |\psi_n\rangle \in S_{\rm tot}=1 \\
\end{array}
\right. , 
\end{eqnarray}
where $|\psi_n\rangle \in S_{\rm tot}=1$ ($|\psi_n\rangle \not\in S_{\rm tot}=1$) indicates 
that $|\psi_n\rangle$ belongs (does not belong) to the $S_{\rm tot}=1$ subspace. 
Therefore, the different behavior of the local spin density under the applied pinning magnetic field 
should be attributed to the amount of the low-lying triplet excitations. 

\begin{figure}[htbp]
\begin{center}
\includegraphics[width=\hsize]{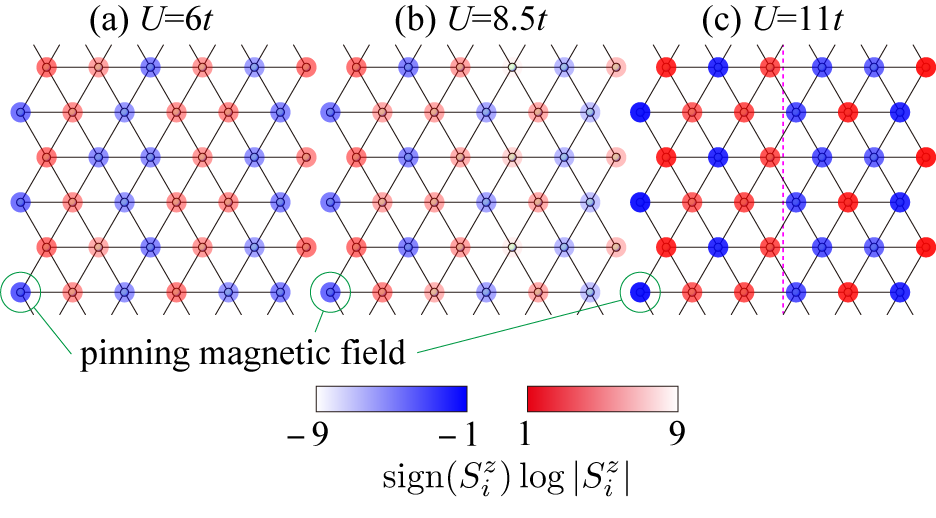}
\caption{(color online) 
Same as Figs.~\ref{pinning}(a)--\ref{pinning}(c) but in the logarithmic scale. 
}
\label{fig:pinning_logplot}
\end{center}
\end{figure}

Figure~\ref{fig:pinning_logplot} shows the same result as in Fig.~\ref{pinning} but in the logarithmic scale. 
It clearly shows that the local spin density $S_i^z$ for $U=8.5t$ in phase II exhibits 
comparable amplitude with that for $U=6t$ in phase I where the ground state is the 
paramagnetic metal 
and thus no triplet excitation gap is expected.
On the other hand, the local spin density $S_i^z$ for $U=8.5t$ 
is significantly smaller than that for $U=11t$ in phase III where the ground state is 
$120^{\circ}$ N\'eel ordered. 
The similarity to the $U=6t$ case thus indicates that there exist an extensive 
amount of gapless spin excitations in phase II 
expected in the thermodynamic limit.

\subsection{\label{sec:spincorrelation}Spin correlation function}

\begin{figure}[htbp]
\begin{center}
\includegraphics[width=\hsize]{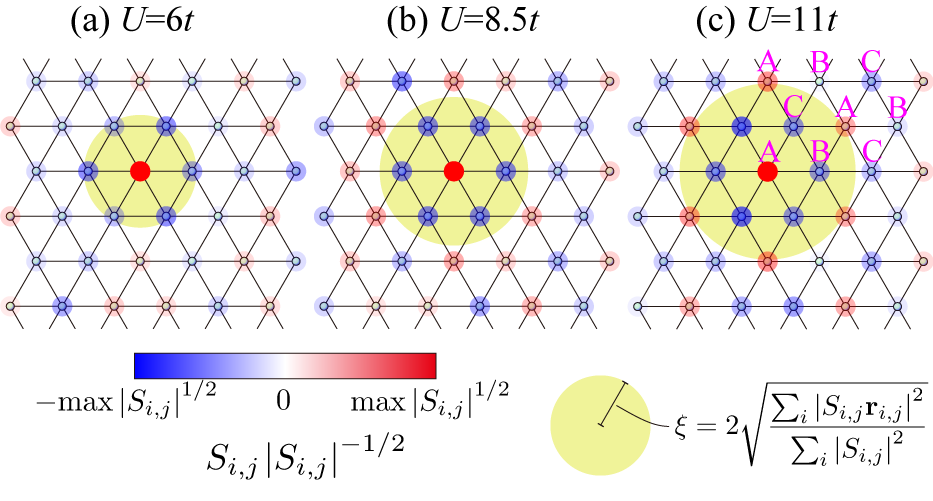}
\caption{(color online) Spin correlation function $S_{i,j} /\sqrt{\left| S_{i,j} \right|}$ 
between a reference site $j$ at the center of the cluster (indicated by a red circle) and other sites $i$ 
for (a) $U=6t$, (b) $U=8.5t$, and (c) $U=11t$. 
Here, $\max|S_{i,j}|^{1/2}$ represents the maximum value of $|S_{i,j}|^{1/2}$ for each $U$. 
Yellow-green shaded circle indicates the correlation length $\xi$.  
A three sublattice pattern expected for the 120$^{\circ}$ N\'eel order is 
indicated in (c) by ``A,'' ``B,'' and ``C,'' where 
$S_{i,j}>0$ ($S_{i,j}<0$) if sites $i$ and $j$ belong to the same (different) sublattice. 
The cluster used here is the 36-site cluster shown in Fig.~\ref{cylinder}(a). 
}
\label{sscor}
\end{center}
\end{figure}

Next, we calculate the spin correlation 
\begin{equation}
S_{i,j} = \langle \psi_0 | \mathcal{S}_i^z \mathcal{S}_j^z | \psi_0 \rangle
\end{equation}
between a reference site $j$ located at the center of the cluster and other sites $i$. 
The representative results for the three different phases are shown in Fig.~\ref{sscor}.
Figure~\ref{sscor}(c) clearly shows that $S_{i,j}$ for $U=11t$ in phase III exhibits a three sublattice pattern, 
compatible with the $120^{\circ}$ N\'eel order. On the other hand, $S_{i,j}$ in phases I and II does not 
show such a three sublattice pattern [see Figs.~\ref{sscor}(a) and \ref{sscor}(b)], strongly suggesting that 
these phases are not $120^{\circ}$ N\'eel ordered.

We notice, however, in Fig.~\ref{sscor} that the intensity of $S_{i,j}$ 
between distant sites for $U=8.5t$ is comparable to that for $U=11t$, 
indicating the relatively long spin correlation in phase II. 
For a quantitative comparison, we estimate the correlation length $\xi$ 
via 
\begin{equation}
\xi=2\sqrt{\frac{\sum_{i=1}^N|S_{i,j}{\bf r}_{i,j}|^2}{\sum_{i=1}^N|S_{i,j}|^2}},
\end{equation}
where ${\bf r}_{i,j} = {\bf r}_i - {\bf r}_j$ and ${\bf r}_i$ is the position vector of site $i$. 
As shown in Fig.~\ref{sscor}, we find that $\xi$ for $U=8.5t$ is shorter than that for $U=11t$ 
where $\xi$ diverges in the thermodynamic limit, 
but is longer than that for $U=6t$ where the spin correlation decays algebraically. 
The rather long range spin correlation in phase II should be contrasted with the exponential decay of spin 
correlation expected in gapped QSL. 
Recalling also the result of the response to the pinning magnetic field, 
the ground state in phase II resembles gapless QSL.

\subsection{\label{sec:sq}Spin structure factor}

The spin structure factor $S({\bf q})$ is the Fourier transform of the real-space spin correlation function 
$S_{i,j}$ and defined as 
\begin{eqnarray}
S({\bf q}) = \sum_{i=1}^N S_{i,j} e^{{\rm i}{\bf q}\cdot ({\bf r}_i - {\bf r}_j)}, 
\end{eqnarray}
where 
site $j$ is a representative site 
chosen at the central site of the cluster as in Fig.~\ref{sscor}. 
Although the wave number ${\bf q}$ is not a good quantum 
number due to open boundary conditions in the $x$ direction, 
here we calculate $S({\bf q})$ for arbitrary ${\bf q}$.

\begin{figure}[htbp]
\begin{center}
\includegraphics[width=\hsize]{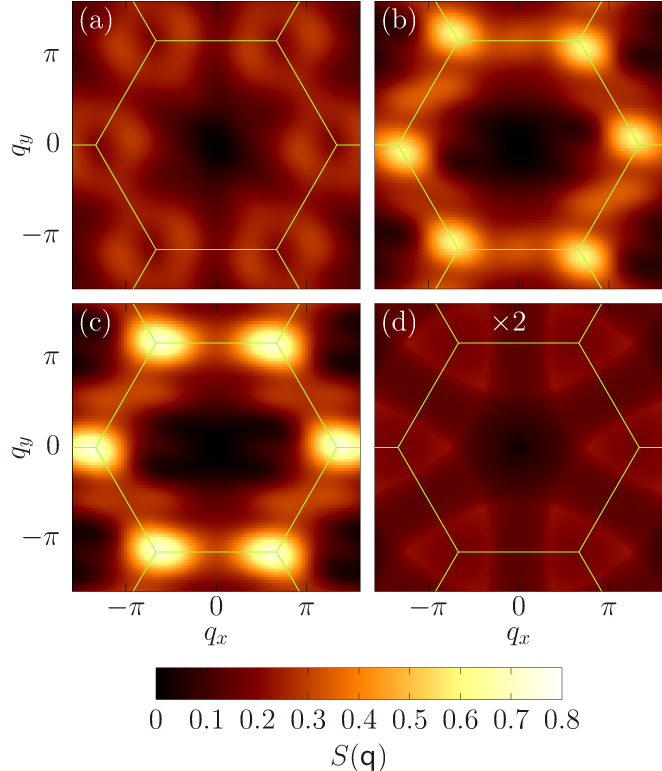}
\caption{(color online) 
Intensity plot of spin structure factor $S({\bf q})$ for (a) $U=6t$, (b) $U=8.5t$, and (c) $U=11t$ 
calculated in the 36-site cluster and for (d) $U=3t$ obtained by the random 
phase approximation. 
Notice that the intensity is doubled in (d) for clarity. 
The Brillouin zone boundaries are indicated by yellow-green lines. 
}
\label{fig:sq}
\end{center}
\end{figure}

The representative results for the three different phases are shown in Figs.~\ref{fig:sq}(a)--\ref{fig:sq}(c). 
We find in Fig.~\ref{fig:sq}(c) that $S({\bf q})$ for $U=11t$ in phase III displays sharp peaks at 
${\bf q} = (2\pi/3,2\pi/\sqrt{3})$ (the ${\rm K}$ point) and other equivalent ${\bf q}$'s 
including the ${\rm K}^{\prime}$ point at ${\bm q} = (-2\pi/3,2\pi\sqrt{3})$, 
which is compatible with the $120^{\circ}$ N\'eel-ordered state. 
The $S({\bf q})$ for $U=8.5t$ in phase II shown in Fig.~\ref{fig:sq}(b) also exhibits broad maxima at the ${\rm K}$ point 
and other equivalent ${\bf q}$'s, but the peak structure is softened as compared with $S({\bf q})$ 
for $U=11t$. 
In contrast, we find in Fig.~\ref{fig:sq}(a) that $S({\bf q})$ for $U=6t$ in phase I shows 
enhanced intensities forming a ring-like structure around the K point (and other 
equivalent ${\bf q}$'s),  
which implies the presence of $2{\bf k}_{\rm F}$ scattering, where ${\bf k}_{\rm F}$ is the Fermi momentum in the 
noninteracting limit.

In order to better understand $S({\bf q})$ in phases I and II, 
we calculate $S({\bf q})$ within the random phase approximation (RPA). 
In the RPA, the spin susceptibility $\chi ({\bf q},z)$ at zero temperature is given as
\begin{eqnarray}
\chi ({\bf q},z) = \frac{\chi_0 ({\bf q},z)}{1 - U \chi_0 ({\bf q},z)}, 
\label{eq:rpa}
\end{eqnarray}
where $z$ is the complex frequency. 
The susceptibility $\chi_0 ({\bf q},z)$ in the noninteracting limit is  
\begin{eqnarray}
\chi_0 ({\bf q},z) = 
\frac{1}{N} \sum_{{\bf k}\in {\rm 1st\,BZ}} \frac{\Theta (-\varepsilon_{\bf k}) - 
\Theta (-\varepsilon_{{\bf k}+{\bf q}})}{z - \varepsilon_{\bf k} + \varepsilon_{{\bf k}+{\bf q}}}, 
\end{eqnarray}
where the sum is taken over the first Brillouin zone (BZ) of the triangular lattice, 
$\Theta (x)$ is the Heaviside step function, 
and $\varepsilon_{\bf k}$ is the noninteracting band dispersion 
\begin{eqnarray}
\varepsilon_{\bf k} &=& - 2 t \cos k_x - 2 t \cos \left( \frac{k_x}{2} + \frac{\sqrt{3}k_y}{2} \right) \nonumber \\
&{}& - 2 t \cos \left( - \frac{k_x}{2} + \frac{\sqrt{3}k_y}{2} \right) - \mu. 
\end{eqnarray}
The chemical potential $\mu$ is tuned such that the electron density is $0.5$ per spin. 
We set the chemical potential $\mu \sim 0.8347t$ for the calculation in the thermodynamic limit.

From $\chi({\bf q},z)$ obtained above within the RPA, 
the spin structure factor $S({\bf q})$ is evaluated as 
\begin{eqnarray}
S({\bf q}) = \frac{1}{\pi} \int_0^{\infty} {\rm d}x {\rm Re} \chi ({\bf q},{\rm i}x). 
\end{eqnarray}
In deriving the above equation, we have assumed that the $z$ component of total spin 
is zero and the system is invariant under the global spin flip. 
Figure~\ref{fig:sq}(d) shows $S({\bf q})$ for $U=3t$ within the RPA. 
Here we choose relatively small $U$ because $\chi({\bf q},0)$ diverges at $U=3.7$--$3.8t$. 
As shown in Fig.~\ref{fig:sq}(d), $S({\bf q})$ exhibits 
a triangular shell-like structure around the K point (and other equivalent ${\bf q}$'s). 
The ridges of the shells lie exactly along the $2{\bf k}_{\rm F}$ lines and the local minimum in the 
center of the shell is located at the K point (and other equivalent ${\bf q}$'s). These features are 
indeed similar to those found in Fig.~\ref{fig:sq}(a) for $U=6t$.

The spinon Fermi sea state is a kind of gappless spin liquid state and has been considered as a candidate 
for the ground state of triangular lattice systems~\cite{motrunich}.
Due to the presence of spinon Fermi surface, $S({\bf q})$ for the spinon Fermi sea state exhibits singularities 
along the $2{\bf k}_{\rm F}$ lines~\cite{block} and is expected to be 
similar to those shown in Figs.~\ref{fig:sq}(a) and \ref{fig:sq}(d). 
However, as shown in Fig.~\ref{fig:sq}(b), we find that 
$S({\bf q})$ for $U=8.5t$ in phase II is quite different from those in Figs.~\ref{fig:sq}(a) and \ref{fig:sq}(d). 
Therefore, the spinon Fermi sea state is unlikely to be the ground state in phase II.

The similarity of $S({\bf q})$ for $U=8.5t$ and $U=11t$ in Fig.~\ref{fig:sq} tempts us to 
conclude that also the ground state in phase II shows tendency towards the $120^{\circ}$ N\'eel order. 
However, we emphasize that the peak structure in $S({\bf q})$ for 
$U=8.5t$ in phase II are much smaller and broader than that for $U=11t$ in phase III. 
Indeed, as shown in Fig.~\ref{pinning} and Fig.~\ref{sscor}, 
the response to the pinning magnetic field and 
the real-space spin correlation function are clearly different in phases II and III.

\subsection{\label{sec:nn_spin}Nearest-neighbor spin correlation}

\begin{figure}[tbhp]
\begin{center}
\includegraphics[width=\hsize]{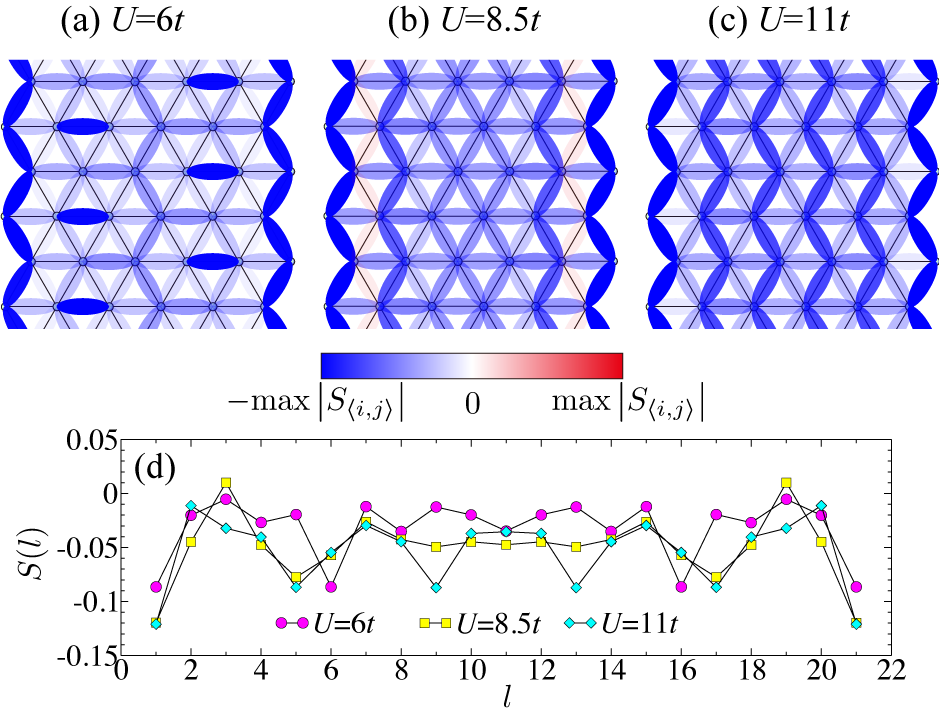}
\caption{(color online) Nearest-neighbor spin correlation $S_{\left< i,j\right>}$ 
for (a) $U=6t$, (b) $U=8.5t$, and (c) $U=11t$. 
Here, $\max|S_{\left< i,j \right>}|$ represents the maximum value of $|S_{\left< i,j\right>}|$ for each $U$. 
(d) $S(l)=S_{\left< i,j \right>}$ along $x$-direction, 
where the bond index $l$ connecting neighboring sites $i$ and $j$ is shown in Fig.~\ref{cylinder}(a).
The 36-site cluster is used for all figures. 
}
\label{nnspin}
\end{center}
\end{figure}

We calculate the nearest-neighbor spin correlation $S_{\left< i,j\right>}\,(=S_{i,j})$ 
for all nearest-neighbor sites $i$ and $j$, and the representative results for the three 
different phases are shown in Fig.~\ref{nnspin}. 
We first notice in Figs.~\ref{nnspin}(a)--\ref{nnspin}(c) that the results are invariant under the 
translation along the $y$ direction, 
the reflection about mirror planes perpendicular to the $y$ direction, 
and the $180^{\circ}$ rotation around the center of the cluster, 
thus implying that the convergence of our results is satisfactory. 
For better quantitative comparison, we show in Fig.~\ref{nnspin}(d) $S(l)=S_{\left< i,j \right>}$ 
along the $x$ direction, where the bond index $l$ connecting sites $i$ and $j$ is 
denoted in Fig.~\ref{cylinder}(a).

It is clearly observed in Fig.~\ref{nnspin}(d) that $S(l)$ for $U=11t$ in phase III 
is rather enhanced at $l=4n-3$ and 
suppressed at $l=4n-1$ for $n=1, 2, 3, \cdots$. More interestingly, 
the oscillations in $S(l)$ for $U=8.5t$ in phase II are smallest, especially around the center 
of the cluster. 
Note that spatial variation of $S(l)$ is an indication of spatial symmetry breaking, 
e.g., valence bond solid, or a tendency for it, and that in our case the spatial variation is 
induced by open boundary conditions in the $x$ direction. 
The strong suppression of oscillations in phase II is therefore a strong indication of 
the absence of valence bond solid and also 
other possible spatial symmetry breakings.

Let us now compare the nature of the ground state in phase II with 
the $Z_2$ spin liquid ground state of the spin-1/2 
antiferromagnetic Heisenberg model on the triangular lattice with the 
next nearest-neighbor exchange interaction, recently reported in Refs.~\onlinecite{zhu,hu}. 
The cluster used here (Fig.~\ref{cylinder}) 
is an odd cylinder~\cite{zhu,hu} with odd number of sites in the one dimensional unit cell. 
Therefore, according to the Lieb-Schultz-Mattis theorem~\cite{lieb}, the ground state for this cluster 
is degenerate if the excitation gap is finite as in the case for the $Z_2$ spin liquid. 
Reflecting this degeneracy, 
the spatial distribution of the nearest-neighbor spin correlation $S_{\left< i,j \right>}$ exhibits 
the strong alternating oscillation, 
induced by open boundary conditions along the $x$ direction~\cite{zhu,hu}. 
Although the similar oscillation pattern is found in phase III and around the edge of the cluster in 
phase II, the central region of the cluster in phase II does not show such feature (see Fig.~\ref{nnspin}).

\subsection{\label{sec:bond}Nearest-neighbor bond correlation}

\begin{figure}[tbhp]
\begin{center}
\includegraphics[width=0.95\hsize]{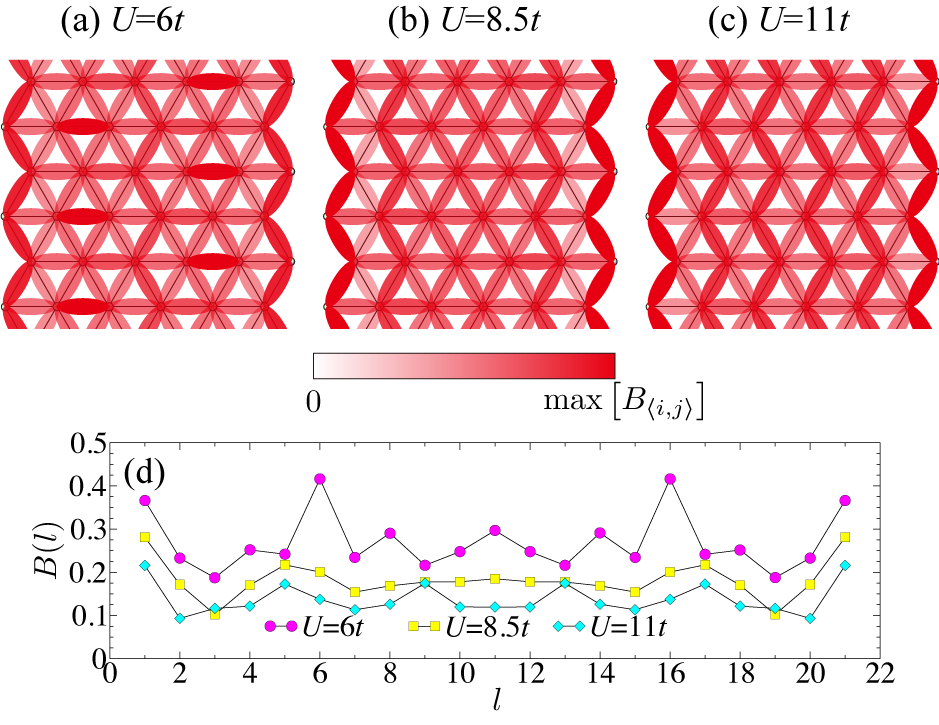}
\caption{(color online) Nearest-neighbor hopping amplitude $B_{\left< i,j\right>}$ for 
(a) $U=6t$, (b) $U=8.5t$, and (c) $U=11t$. 
Here, $\max[B_{\left< i,j\right> }]$ is the maximum value of $B_{\left< i,j \right>}$ for each $U$. 
(d) $B(l)=B_{\left< i,j \right>}$ along the $x$ direction, 
where the bond index $l$ connecting neighboring sites $i$ and $j$ is denoted in Fig.~\ref{cylinder}(a). 
Notice that the increase of kinetic energy, proportional to $-t\sum_l B(l)$,  
is nicely observed with increasing $U/t$.
The 36-site cluster is used for all figures. 
}
\label{nnbond}
\end{center}
\end{figure}

We also calculate the hopping amplitude $B_{\left< i,j\right>}$ between nearest-neighbor sites $i$ and $j$ 
defined as 
\begin{eqnarray}
B_{\left< i,j\right>} = \frac{1}{2}\left< \psi_0 \right| (c_{i,\sigma}^{\dagger} c_{j,\sigma}+c_{j,\sigma}^{\dagger}c_{i,\sigma}) \left| \psi_0 \right>. 
\end{eqnarray}
The representative results for the three different phases are shown in Fig.~\ref{nnbond}. 
Similarly to $S_{\left< i,j \right>}$, 
we find that $B_{\left< i,j \right>}$ is invariant under the translation, reflection, 
and rotation operations [see Figs.~\ref{nnbond}(a)--\ref{nnbond}(c)], 
suggesting that our results are well converged. 
For better quantitative comparison, 
Fig.~\ref{nnbond}(d) shows $B(l)=B_{\left< i,j \right>}$ along the $x$ direction for the $l$th bond 
connecting sites $i$ and $j$ [for the indexing of bonds, see Fig.~\ref{cylinder}(a)].

We find that $B_{\left< i,j \right>}$ exhibits the similar oscillation patterns to those observed in 
$S_{\left< i,j \right>}$ (see Fig.~\ref{nnspin}).
The similarity between $B_{\left< i,j \right>}$ and $S_{\left< i,j \right>}$ is expected for large $U/t$ 
since the kinetic energy, 
proportional to $B_{\left< i,j \right>}$, can be in a strong coupling regime transferred to the Heisenberg exchange interaction, 
which is related to $S_{\left< i,j \right>}$. However, it is surprising that this similarity is present also in the coupling regimes 
shown in Fig.~\ref{nnbond}, where there is in general no direct connection between 
$B_{\left< i,j \right>}$ and $S_{\left< i, j \right>}$. 
As shown in Fig.~\ref{nnbond}(d), we find that the oscillations of $B(l)$ around the center of the cluster 
are most strongly reduced for $U=8.5t$ in phase II as compared with those in phases I and III. 
This implies that the ground state in phase II is not compatible with the nearest-neighbor 
valence bond solid.

\subsection{\label{sec:chiral}Chiral correlation function}

\begin{figure}[tbhp]
\begin{center}
\includegraphics[width=\hsize]{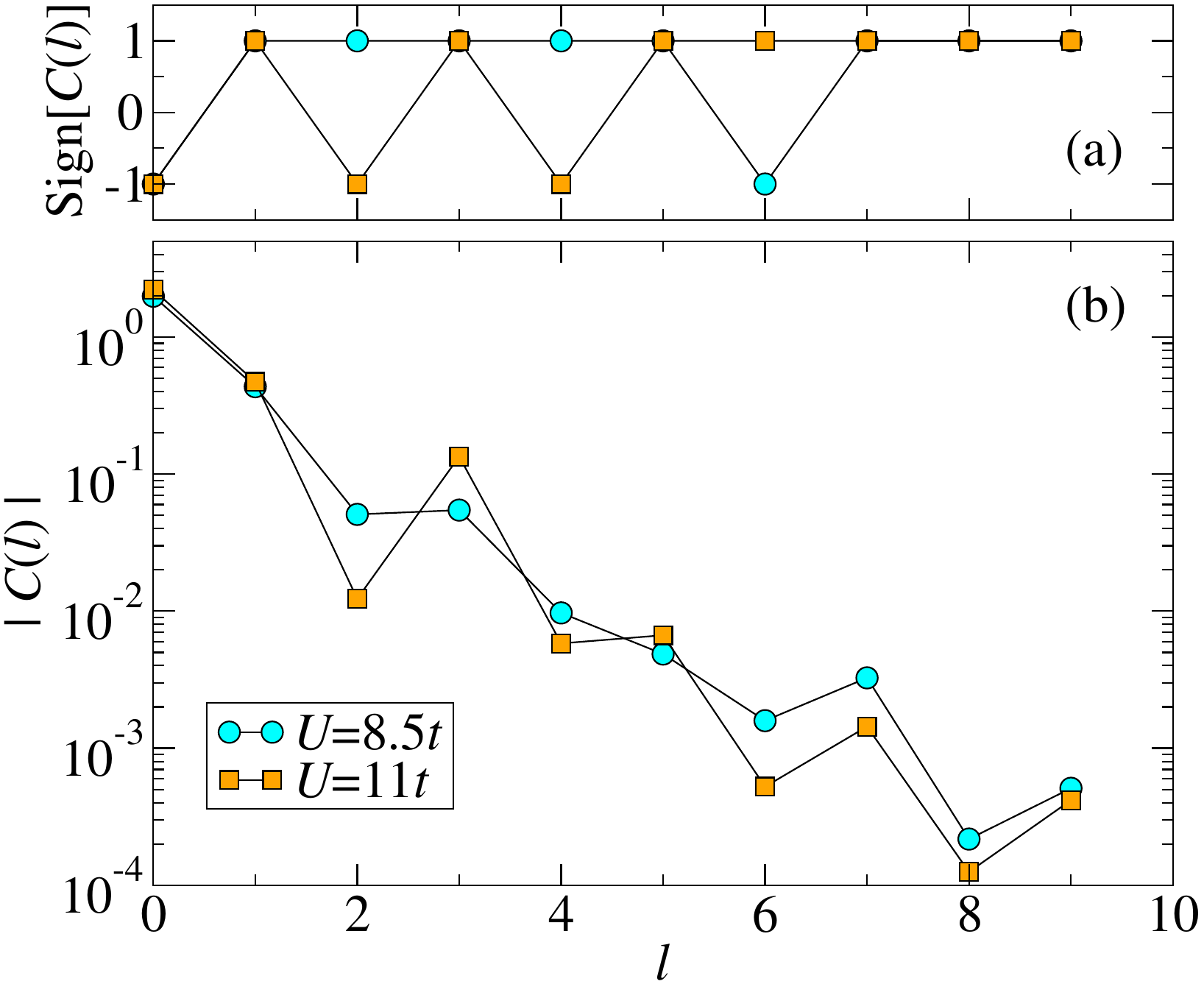}
\caption{(color online) (a) Sign and (b) amplitude of chiral correlation function $C(l)$ for 
$U=8.5t$ and $11t$. 
The cluster used here is the 36-site cluster shown in Fig.~\ref{cylinder}(a). 
}
\label{chiral}
\end{center}
\end{figure}

Next, we calculate the chiral correlation function $C({\triangle_i},{\triangle_j})$ defined as
\begin{eqnarray}
C ({\triangle_i},{\triangle_j}) = \left<\psi_0| C_{\triangle_i} C_{\triangle_j} |\psi_0\right> 
\end{eqnarray}
with
\begin{eqnarray}
C_{\triangle_i} = \vec{\mathcal{S}}_{i_1} \cdot ( \vec{\mathcal{S}}_{i_2} \times \vec{\mathcal{S}}_{i_3} ), 
\end{eqnarray}
where $\triangle_i$ indicates the $i$th elementary triangle 
formed by three neighboring sites $i_1$, $i_2$, and $i_3$ 
in clockwise or counter clockwise order, and the indexing of elementary triangles 
as well as their chiral directions 
is indicated in Fig.~\ref{cylinder}(a). 
$\vec{S}_i$ is the spin operator at site $i$ defined as 
\begin{eqnarray}
\vec{\mathcal{S}}_i = \frac{1}{2} \sum_{\sigma_1=\uparrow,\downarrow} \sum_{\sigma_2 = \uparrow,\downarrow} c_{i,\sigma_1}^{\dagger} \vec{\sigma}_{\sigma_1,\sigma_2} c_{i,\sigma_2}, 
\end{eqnarray}
where $\vec{\sigma} = (\sigma_x, \sigma_y, \sigma_z) $ are Pauli matrices. 
Figure~\ref{chiral} shows the chiral correlation function $C ({\triangle_i},{\triangle_i}+l)=C(l)$ 
with ${\triangle_i}=0$ along the $x$ direction [see Fig.~\ref{cylinder}(a)]. 
We find that the sign of $C (l)$ exhibits nontrivial oscillation [Fig~\ref{chiral}(a)] and 
the amplitude of $C (l)$ decays exponentially [Fig~\ref{chiral}(b)]. 
Therefore, we conclude that the chiral spin liquid is most unlikely to be 
the ground state in phase II~\cite{gong2}.

\subsection{\label{sec:scpair}Pairing correlation function}

\begin{figure}
\includegraphics[width=\hsize]{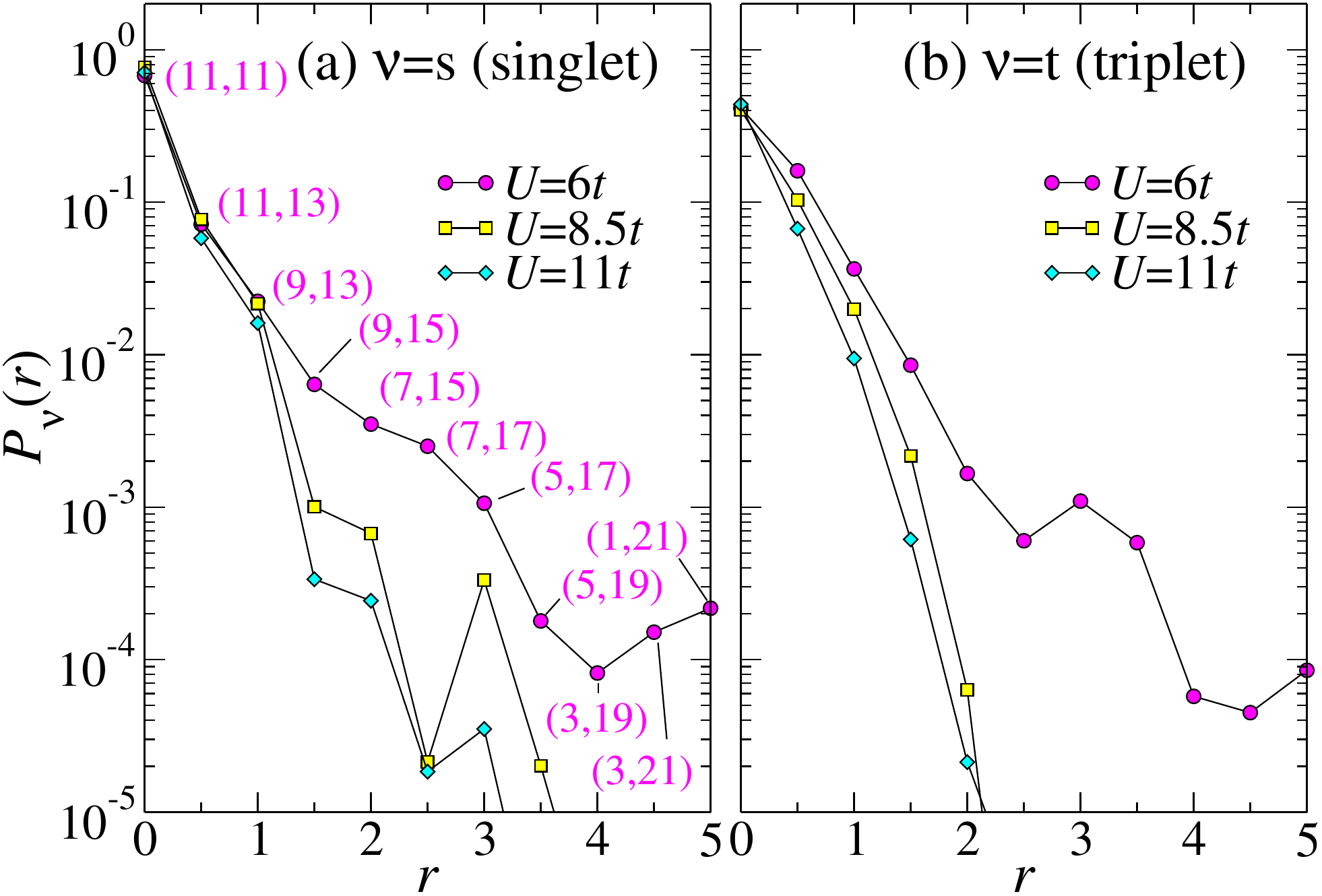}
\caption{(color online) (a) Singlet ($\nu={\rm s}$) and (b) triplet ($\nu={\rm t}$) 
pairing correlation function $P_{\nu}(r)=P_{\nu}(i,j,k,l) $ for $U=6t$, $8.5t$, and $11t$ 
calculated in the 36-site cluster. 
A pair of integer numbers $(l_1,l_2)$ beside each data point for $U=6t$ denotes 
a pair of bond indexes $l_1$ and $l_2$ [for the definition, see Fig.~\ref{cylinder}(a)], 
representing nearest-neighbor sites $(i,j)$ and $(k,l)$, respectively, for which 
the pairing operators $\Delta_\nu (i,j)$ 
and $\Delta^\dag_\nu (k,l)$ are chosen in Eq.~(\ref{eq:scpair}). 
$r$ is the distance between the centers of bonds $l_1$ and $l_2$. 
}
\label{fig:scpair}
\end{figure}

Finally, let us discuss the possibility of superconductivity 
by calculating the pairing correlation function $P_{\nu}(i,j,k,l)$ 
defined as
\begin{equation}
P_{\nu}(i,j,k,l) = \langle \psi_0 \vert \Delta_{\nu}(i,j) \Delta^{\dagger}_{\nu}(k,l) \vert \psi_0 \rangle 
\label{eq:scpair}
\end{equation}
with the nearest-neighbor singlet channel ($\nu = {\rm s}$)
\begin{equation}
\Delta_{\rm s}(i,j) = \frac{1}{\sqrt{2}} \left( c_{i, \uparrow} c_{j, \downarrow} - c_{i, \downarrow} c_{j, \uparrow} \right)
\end{equation}
and the nearest-neighbor triplet channel ($\nu = {\rm t}$)
\begin{equation}
\Delta_{\rm t}(i,j) = \frac{1}{\sqrt{2}} \left( c_{i, \uparrow} c_{j, \downarrow} + c_{i, \downarrow} c_{j, \uparrow} \right).
\end{equation}
Figure~\ref{fig:scpair} shows the representative results of the pairing correlation function 
$P_{\nu}(r) = P_{\nu}(i,j,k,l)$ for both singlet and triplet channels, 
where $r$ is the distance between the centers of two pairs of nearest-neighbor sites $(i,j)$ and $(k,l)$. 
We find that the pairing correlations in both channels are significantly suppressed 
for $U=8.5t$ in phase II and $U=11t$ in phase III as compared with those for $U=6t$ in phase I. 
Therefore, we conclude that the ground-state in phase II is unlikely to be superconducting. 
It should also be noted that the short range superconducting correlations in the 
spin triplet channel is stronger than those in the spin singlet channel for the three representative cases, 
although the superconducting correlations in the spin singlet channel dominates in the long distance.  

It is also interesting to notice that the pairing correlations at long distances seem to be saturated 
for $U=6t$ in phase I.  
However, since the longest distances are calculated from sites 
close to the cluster edges, the upturn of the pairing correlations might be a finite size effect. 
Therefore, our calculations alone cannot support the presence of the superconducting phase. 
Larger systems with $U$ closer to $U_{\rm c1}$ might show stronger pairing correlations. 
This issue is left for the future study.

\section{\label{sec:discussion}Discussion}

Let us briefly discuss implications of our results for the experiments. 
Our results show a rather small discontinuity $\Delta n_d$ ($\sim0.007$) in the double 
occupancy found at the metal-insulator transition $U_{\rm c1}$, 
which is in sharp contrast to the previous studies using other approaches~\cite{yoshioka}, 
where $\Delta n_d$ is typically much larger (from 0.02 to 0.06), 
and is even qualitatively different from the continuous transition 
discussed in Refs.~\onlinecite{senthil} and \onlinecite{mishmash}. 
Smallness of the jump might be the origin of the controversy 
about the order of the transition. 
We note, however, that the discontinuity $\Delta n_d$ corresponds 
to the interaction (or equivalently kinetic)
energy jump of $U_{\rm c1}\Delta n_d \sim 0.05t$, giving $\sim 25$ K for the organic materials with 
$t \sim 50$ meV~\cite{kokalj}, which nicely compares with the temperature where 
the first-order transition line ends in the phase diagram of the organic 
materials~\cite{furukawa}.

Next, let us discuss the ground-state in phase II in terms of RVB states 
described by Gutzwiller projected fermionic wave functions. 
For this purpose, it is important to recall that 
the structure factor $S({\bm q})$ in phase II 
exhibits the maximum at the ${\rm K}$ and ${\rm K}^{\prime}$ points. 
This feature is not consistent with a projected Fermi sea with a large Fermi 
surface because the $2{\bf k}_{\rm F}$ structure is not found in $S({\bm q})$. 
Instead, this feature is rather comparable to a projected 
Fermi sea with gapless nodal points such as a projected Dirac 
fermion~\cite{sindzingre,yunoki2004}.

Another interesting feature is that the superconducting fluctuations for the spin-triplet (singlet) channel 
dominates in the short (long) distance,  although the superconducting correlation functions for both 
channels decay exponentially in the insulating phases.  
This suggests that the long wavelength behavior of the ground-state in phase II might be 
captured by a projected BCS wave function with a singlet pairing, but the strong modification of the 
wave function would be required to describe the short-range properties such as the ground-state energy.

\section{\label{sec:summary}Summary}

In summary, we have performed large scale 2D-DRMG calculations, 
using up to $m=20\ 000$ density-matrix eigenstates, 
to examine the ground state phase diagram of the Hubbard model 
on the triangular lattice at half-filling. 
We have shown that the convergence of our results is well controlled  
and, therefore, our results can be regarded as the most accurate 
and unbiased results available at a moment, apart from the small cluster shape and size dependence.

We have found two first-order transitions separating the three different phases,
which include the metallic phase in weak coupling region, the 120$^{\circ}$ N\'eel-ordered phase 
in strong coupling region, and the QSL like phase in the intermediate couplin region. 
The weak and intermediate coupling phases are less affected by the pinning magnetic fields, 
suggesting the absence of magnetic order in these two phases. 
The spin correlations in the intermediate phase is weaker than those 
in the 120$^{\circ}$ N\'eel-ordered phase and stronger than those in the metallic phase. 
The spin structure factor in the intermediate phase shows a maximum at the ${\rm K}$ and ${\rm K}^{\prime}$ points, 
which is not compatible with the spinon Fermi sea state~\cite{motrunich}. 
The spatial distribution of the nearest-neighbor spin correlation in the intermediate phase 
is not comparable with the $Z_2$ spin liquid found in the spin-1/2 antiferromagnetic Heisenberg 
model on the triangular lattice with the next-nearest-neighbor exchange interaction~\cite{zhu,hu}. 
We have also calculated the chiral correlation function and found that 
the chiral spin liquid~\cite{gong2} is unlikely in the intermediate phase. 
The pairing correlation function decreases monotonically with 
increasing $U/t$, suggesting that the superconductivity is also unlikely 
in the intermediate phase.

The clusters used here are much smaller than those employed for the 2D-DMRG studies of 
spin-1/2 antiferromagnetic Heisenberg models on the triangular lattice reported in Refs.~\onlinecite{zhu,hu}. 
This is simply because the local degrees of freedom in the Hubbard model is two times larger than 
those in the spin-1/2 Heisenberg models. 
Therefore, the more detail analysis using larger clusters is highly 
desirable in order to determine the nature of the ground state 
in the intermediate phase and, in particular, to address the size of the spin gap in the thermodynamic limit 
and the experimental observation in EtMe$_3$Sb[Pd(dmit)$_2$]$_2$ where gapless QSL 
is suggested~\cite{yamashita1}. 
Further properties of the intermediate phase, including the nature of excitations, 
remain to be firmly examined since that would greatly improve our
understanding of spin liquid in general as well as of the organic materials in particular.

\acknowledgments

The authors are grateful to S. Nishimoto and T. Li for valuable discussion. 
This work has been supported by Grant-in-Aid for Scientific Research from the Japan Society for 
the Promotion of Science (JSPS) (Grant No. 24740269, No. 26287079, and No. 17K14148) and 
Slovenian Research Agency (Z1-5442), 
and in part by RIKEN Molecular Systems and RIKEN iTHES Project. 
T. T. and J. K. acknowledge the visiting program for young researchers 
at Yukawa Institute for Theoretical Physics, Kyoto University. 
The computation has been performed using the RIKEN Cluster of Clusters (RICC), the 
RIKEN supercomputer system (HOKUSAI GreatWave), and the K computer at RIKEN 
Advanced Institute for Computational Science (AICS) under 
MEXT HPCI Strategic Programs for Innovative Research (SPIRE) 
(Project Nos. hp120137, hp140128, hp150112, hp160122, and hp170324).

\appendix

\section{\label{app:egap}Entanglement Gap}

In this appendix, we examine the charge and spin gaps 
in the low-lying entanglement spectrum of the ground state~\cite{lihaldane}
to support our results in the main text. 
In the DMRG method, the system is divided into two regions,
blocks $A$ and $B$, and thus 
the ground state $\vert \psi \rangle$ is represented as 
\begin{equation}
\vert \psi \rangle = \sum_{i,j} \psi_{ij} \vert i \rangle_A \vert j \rangle_B,
\end{equation}
where $\vert i \rangle_A$ ($\vert j \rangle_B$) denotes a basis in block $A$ ($B$).
The reduced density matrix $\rho_A$ for block $A$ 
is obtained by tracing out the degrees of freedom in block $B$, 
\begin{equation}
\rho_A = {\rm Tr}_B \vert \psi \rangle \langle \psi \vert,
\end{equation}
where ${\rm Tr}_B$ indicates the trace over all bases in block $B$.
The entanglement spectrum $\xi_n$ (where $n=1,2,3,\dots$) is defined as 
\begin{equation}
\xi_n = -\ln \lambda_n, 
\label{eq:def:espec}
\end{equation}
where $\lambda_n$ is the $n$th largest eigenvalue of the reduced density matrix $\rho_A$. 
Since $0<\lambda_n<1$ in general, $\xi_1 \leq \xi_2 \leq \xi_3 \leq\cdots$.

Equation (\ref{eq:def:espec}) implies that $\xi_n$ can be considered as the eigenvalues of 
the entanglement Hamiltonian $H_{\rm E}$ defined as 
\begin{equation}
H_{\rm E} = - \ln \rho_A.
\end{equation}
This in turn suggests that $H_{\rm E}$ can be regarded as an effective Hamiltonian 
to represent the density matrix $\rho_A$ with the Boltzmann distribution ${\rm e}^{-H_{\rm E}}$. 
Since the density matrix can be block diagonalized with respect to the number of electrons $N_e$ 
and the $z$-component $S^z$ of the total spin in block $A$, 
the entanglement spectrum $\xi_n$ is also 
labelled with these quantum numbers, i.e., $\xi_n = \xi(k, N_e, S^z)$, where 
$k\,(=0,1,2,\dots)$ is an index to distinguish the entanglement spectrum in the same 
quantum number sector: 
$\xi(0,N_e,S^z) \leq \xi(1,N_e,S^z) \leq \xi(2,N_e,S^z) \leq \cdots$. 
We can now define the entanglement gaps for the charge sector as 
\begin{align}
\Delta \xi_{\rm C} = {\rm min}[
& \xi(0,N/2+1,1/2) - \xi(0,N/2,0), \nonumber \\
& \xi(0,N/2-1,1/2) - \xi(0,N/2,0)]
\end{align}
and for the spin sector as 
\begin{align}
\Delta \xi_{\rm S} = {\rm min}[
& \xi(0,N/2,1) - \xi(0,N/2,0), \nonumber \\
& \xi(0,N/2,-1)-\xi(0,N/2,0)],
\end{align}
where the size of block $A$ is half of the cluster size $N$.

\begin{figure}[htbp]
\includegraphics[width=\hsize]{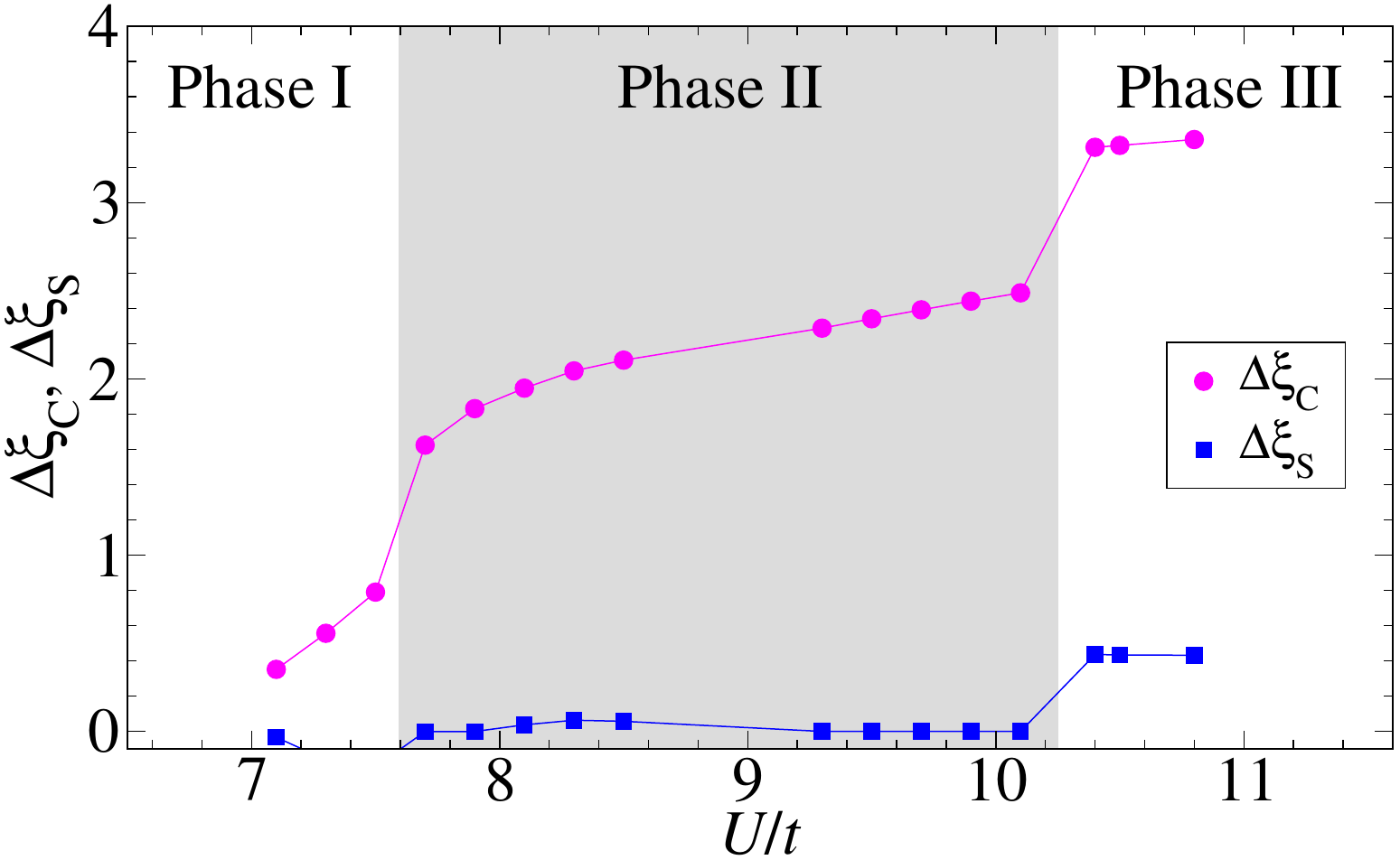}
\caption{(color online) 
Entanglement gaps $\Delta \xi_{\rm C}$ and $\Delta \xi_{\rm S}$ for the charge and spin sectors, respectively. 
A shaded region indicates phase II determined from the discontinuities of the double occupancy 
in the $48$-site cluster. 
}
\label{fig:egaps}
\end{figure}

Figure~\ref{fig:egaps} shows the entanglement gaps $\Delta \xi_{\rm C}$ and $\Delta \xi_{\rm S}$ 
as a function of $U/t$. 
We indeed find that $\Delta \xi_{\rm C}$ increases abruptly at the phase boundaries.
Since the value of $\Delta \xi_{\rm C}$ is related inversely to the 
global charge fluctuations between blocks A and B, the abrupt increase of $\Delta \xi_{\rm C}$ 
in the phase boundary between phases I and II suggests that this transition involves the 
opining of charge gap. 
Therefore, one would regard the transition between phases I and II as the Mott transition. 

We also find that the entanglement gap $\Delta\xi_{\rm S}$ for the spin sector increases 
abruptly at the phase boundary between phases II and III. 
Moreover, we find that the $\Delta \xi_{\rm S} \sim 0$ in the phase II. 
This tempts us to conclude that the ground state is spin gapless in phase II. 
However, this is not appropriate because a topologically non-trivial gapped state 
can induce characteristic low-lying edge states with 
the gapless entanglement spectrum for the spin sector. 
Therefore, we can only argue from Fig.~\ref{fig:egaps} that the spin structure of phase II is distinguishable 
from phase III.


\end{document}